\documentclass[%
aps,
 pra,
 amsmath,amssymb,
 reprint,%
 superscriptaddress,
]{revtex4-1}

\usepackage{dcolumn}
\usepackage{graphicx}
\usepackage{lineno}
\usepackage{bm}%
\usepackage[pdftex,colorlinks=true,bookmarks=false,citecolor=blue,urlcolor=blue]{hyperref}
\usepackage{upgreek}

\usepackage{titlesec} % Allows customization of titles
\begin{document}
%%%%%%%%%%%%%%%%%%%%%%%%%%%%%%%%%%%%%%%%%%%%%%%%%%%%%%%%%%%%%%%%%%%%%
%% TITLE SECTION
%%%%%%%%%%%%%%%%%%%%%%%%%%%%%%%%%%%%%%%%%%%%%%%%%%%%%%%%%%%%%%%%%%%%%

\title{\vspace{-15mm}\fontsize{19pt}{10pt}\selectfont\textbf{Optical parametric generation in a lithium niobate microring with modal phase matching}} % Article title
\author{Rui Luo}
\thanks{These two authors contributed equally.}
\affiliation{Institute of Optics, University of Rochester, Rochester, NY 14627}

\author{Yang He}
\thanks{These two authors contributed equally.}
\affiliation{Department of Electrical and Computer Engineering, University of Rochester, Rochester, NY 14627}

\author{Hanxiao Liang}
\affiliation{Department of Electrical and Computer Engineering, University of Rochester, Rochester, NY 14627}

\author{Mingxiao Li}
\affiliation{Department of Electrical and Computer Engineering, University of Rochester, Rochester, NY 14627}

\author{Jingwei Ling}
\affiliation{Institute of Optics, University of Rochester, Rochester, NY 14627}

\author{Qiang Lin}
\email[Electronic mail: ]{qiang.lin@rochester.edu}
\affiliation{Institute of Optics, University of Rochester, Rochester, NY 14627}
\affiliation{Department of Electrical and Computer Engineering, University of Rochester, Rochester, NY 14627}

\date{\today}

%----------------------------------------------------------------------------------------

%%%%%%%%%%%%%%%%%%%%%%%%%%%%%%%%%%%%%%%%%%%%%%%%%%%%%%%%%%%%%%%%%%%%%
%% ABSTRACT
%%%%%%%%%%%%%%%%%%%%%%%%%%%%%%%%%%%%%%%%%%%%%%%%%%%%%%%%%%%%%%%%%%%%%

\begin{abstract}
	
The lithium niobate integrated photonic platform has recently shown great promise in nonlinear optics on a chip scale. Here, we report second-harmonic generation in a high-$Q$ lithium niobate microring resonator through modal phase matching, with a conversion efficiency of 1,500$\%~\rm{W^{-1}}$. Our device also allows us to observe difference-frequency generation in the telecom band. Our work demonstrates the great potential of the lithium niobate integrated platform for nonlinear wavelength conversion with high efficiencies.

\end{abstract}

\maketitle % Insert title

%%%%%%%%%%%%%%%%%%%%%%%%%%%%%%%%%%%%%%%%%%%%%%%%%%%%%%%%%%%%%%%%%%%%%
%% INTRODUCTION
%%%%%%%%%%%%%%%%%%%%%%%%%%%%%%%%%%%%%%%%%%%%%%%%%%%%%%%%%%%%%%%%%%%%%
\section{Introduction}

Optical parametric generation via a quadratic nonlinearity has been extensively studied for the capability of wavelength conversion through elastic photon-photon scattering, constituting the basis of various applications including coherent radiation \cite{Dunn99}, spectroscopy \cite{Rosenman99}, frequency metrology \cite{Cundiff03}, and quantum information processing \cite{Pan12}. With the ability to strongly confine optical modes in the micro-/nano-scale, a number of integrated photonic platforms have been developed for strong nonlinear optical effects with high efficiencies and low power consumption \cite{Harris06, Vuckovic09, Lipson11, Pavesi12, Solomon14, Tang16Optica, Watts17}.

Among all the integrated nonlinear photonic platforms, lithium niobate (LN) has recently attracted remarkable attentions, owing to its wide transparency window and strong quadratic optical nonlinearity. To date, a variety of nanophotonic systems, including waveguides \cite{Pertsch15, Bowers16, Loncar17OE, Fathpour17APL, Luo18Optica, Ding18, Luo18semi}, microdisks \cite{Loncar14, Cheng16, Luo17OE, Xu17, Cheng18, Xiao18}, microrings \cite{Zappe18, Huang18}, and photonic crystal cavities \cite{Pertsch13, Jiang18, Li18}, have been studied for optical parametric processes in LN. Cavity-enhanced nonlinear wavelength conversion has been demonstrated in doubly/triply resonant LN microresonators through a number of techniques including modal phase matching \cite{Loncar14, Xu17, Xiao18, Huang18}, cyclic phase matching \cite{Cheng16, Luo17OE, Cheng18}, and quasi-phase matching \cite{Zappe18}. However, the potential of the LN integrated platform has not yet been fully explored for efficient nonlinear parametric processes, and current devices demonstrate only moderate efficiencies far from what LN can provide. Here, we report optical parametric generation in a high-$Q$ Z-cut LN microring resonator through exact modal phase matching. The device exhibits optical $Q$'s of $\sim$$10^5$ for the designed cavity modes in the 1550 and 780 nm bands, and both modes are well coupled to a single bus waveguide, enabling us to conveniently measure a second-harmonic generation (SHG) efficiency of 1,500$\%~\rm {W^{-1}}$. In addition, by pumping into the mode in the 780 nm band, we are also able to observe difference-frequency generation (DFG) in the telecom band. Our work shows the great promise of modal-phase-matched LN microresonators for efficient optical parametric generation.

\begin{figure*}[t!]
	\centering\includegraphics[width=2\columnwidth]{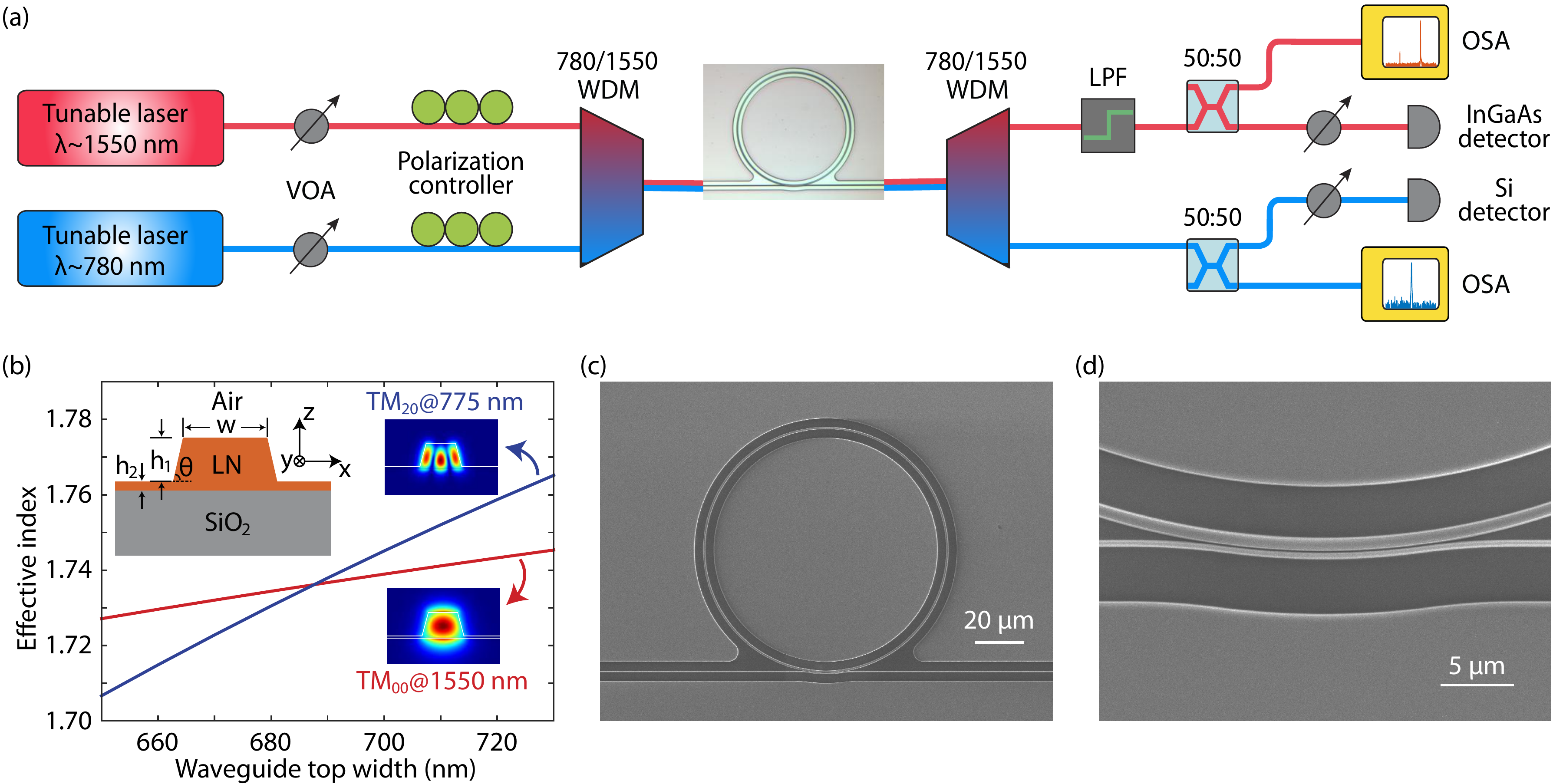}
	\caption{(a) Experimental setup for device characterization and optical parametric generation. VOA: variable optical attenuator; WDM: wavelength-division multiplexer; LPF: long-pass filter; OSA: optical spectrum analyzer. (b) Numerically simulated effective indices of the TM$_{00}$ mode at 1550 nm and the TM$_{20}$ mode at 775 nm, as functions of the top width $w$ of a straight waveguide. Other waveguide parameters are $h_1$=550 nm, $h_2$=50 nm, and $\theta$=75$^\circ$. (c) Scanning electron microscopy image of our LN microring. (d) Zoom-in of the bus-ring coupling region.} \label{Fig1}
\end{figure*}

\section{Design and characterization}

In order to achieve modal phase matching in a microresonator, we performed photonic design with a Z-cut LN thin film, whose optical axis lies vertically, showing no anisotropy of refractive index in the device plane. To utilize the largest nonlinear term $d_{33}$, we designed for phase matching between the fundamental quasi-transverse-magnetic mode (TM$_{00}$) at 1550 nm and a high-order mode TM$_{20}$ at 775 nm. For simplicity, we preformed numerical simulation of effective indices in a straight waveguide, as a guideline for microring resonators with a relatively large radius, which is 50 $\mu$m in our study. Fig.~\ref{Fig1}(b) presents the simulation result by the finite element method, which shows that for a waveguide thickness of 600 nm, modal phase matching happens for TM$_{00}$ at 1550 nm and TM$_{20}$ at 775 nm when the waveguide width is about 690 nm. For a microring resonator with the same cross-section, since the Z-cut LN thin film is isotropic in the device plane, the phase matching condition is consistently satisfied at any azimuthal angle, which is expected to produce strong SHG as the phase-matched FF light travels around the cavity.

Our device fabrication started from a Z-cut LN-on-insulator wafer by NANOLN, with a 600-nm-thick LN thin film sitting on a 3-$\mu$m-thick buried oxide layer and a silicon substrate, and the process was similar to that of our previous work \cite{Li18}. Fig.~\ref{Fig1}(c) shows a fabricated microring resonator, coupled to a pulley waveguide \cite{Adibi10, Lu18}, and Fig.~\ref{Fig1}(d) gives a closer look at the coupling region. Later device characterization shows that a bus waveguide top width of $\sim$200 nm, a gap (measured at the top surface of the LN thin film) of $\sim$350 nm, and a coupling length of $\sim$20 $\mu$m are able to give good coupling for both the fundamental-frequency (FF) and the second-harmonic (SH) modes.

After fabricating the device, we conducted experiments to characterize its linear optical properties and demonstrate nonlinear parametric generation, with the setup shown in Fig.~\ref{Fig1}(a). We used two continuous-wave tunable lasers, one in the telecom band around 1550 nm, the other in the near-infrared (NIR) around 780 nm. Light from both lasers was combined by a 780/1550 wavelength-division multiplexer (WDM), and launched into the on-chip bus waveguide via a lensed fiber. The bus waveguide coupled light at both wavebands into and out of the microring resonator, inside which nonlinear optical parametric processes took place. A second lensed fiber was used to collect output light from the chip, and a second 780/1550 WDM was utilized to separate light at the two wavebands. At the 1550 port of the WDM, a long-pass filter that passes light with a wavelength over 1100 nm was used to eliminate residual NIR light, and the telecom light was further split into two paths, one to an InGaAs detector for characterization, and the other to an optical spectrum analyzer (OSA) for spectral analysis of DFG; at the 780 port, the NIR light was also split into two paths, one to a Si detector for characterization, and the other to an OSA for detection of SHG. Variable optical attenuators were employed to study power-dependent properties, and polarization controllers were used for optimal coupling of the wanted polarization.

\begin{figure}[t!]
	\centering\includegraphics[width=1\columnwidth]{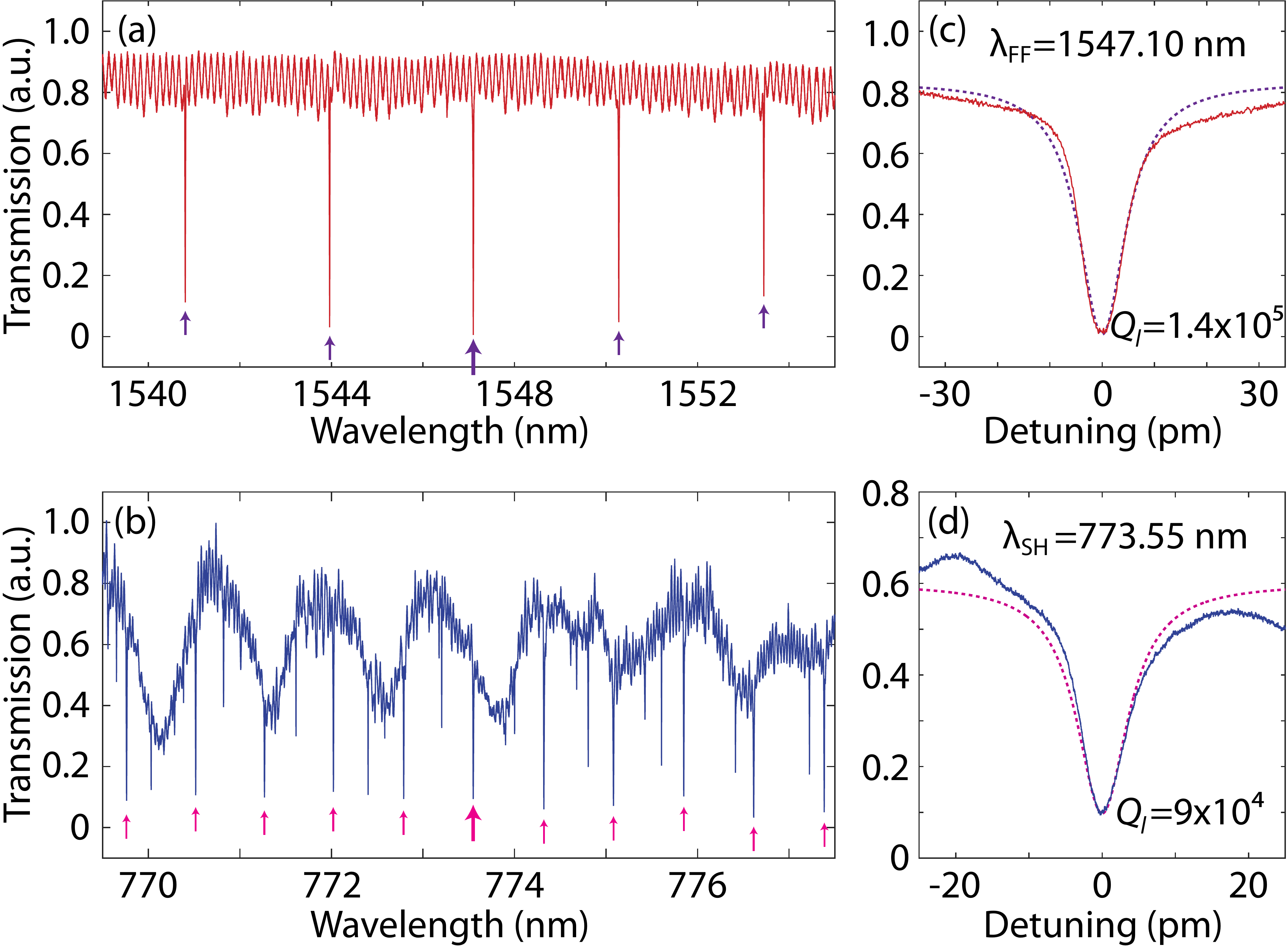}
	\caption{Transmission spectra of the LN microring near (a) 1550 nm, and (b) 780 nm. TM$_{00}$ modes around 1550 nm and TM$_{20}$ modes around 780 nm are indicated by purple and magenta arrows, respectively, with big arrows showing the phase-matched modes. (c) and (d) Detailed transmission spectra of the two phase-matched modes, with experimental data shown as solid curves and fittings shown as dashed curves. } \label{Fig2}
\end{figure}

In order to obtain the linear optical properties of our microring resonator, we scanned the wavelengths of both lasers and measured the transmission spectra near both 1550 and 780 nm, as shown in Fig.~\ref{Fig2}(a) and (b). Our microring resonator exhibits a single TM mode family near 1550 nm, and the mode at 1547.10 nm, which is the FF mode for modal-phase-matched SHG, is almost critically coupled, with a coupling depth of $\sim$99\% and a loaded optical $Q$ of 1.4$\times10^5$ [see Fig.~\ref{Fig2}(c)]. On the other hand, the SH mode at 773.55 nm is under-coupled, with a coupling depth of $\sim$83\% and a loaded optical $Q$ of 9$\times10^4$ [see Fig.~\ref{Fig2}(d)]. The fiber-to-chip coupling losses are about 6.9 and 11.4 dB/facet for the FF and SH modes, respectively. These high optical $Q$'s, together with the large nonlinearity in the designed type-0 process using $d_{33}=27$ pm/V, indicate strong and efficient nonlinear optical interactions in phase-matched parametric generation with cavity enhancement.

\section{Optical parametric generation}

\begin{figure}[t!]
	\centering\includegraphics[width=0.9\columnwidth]{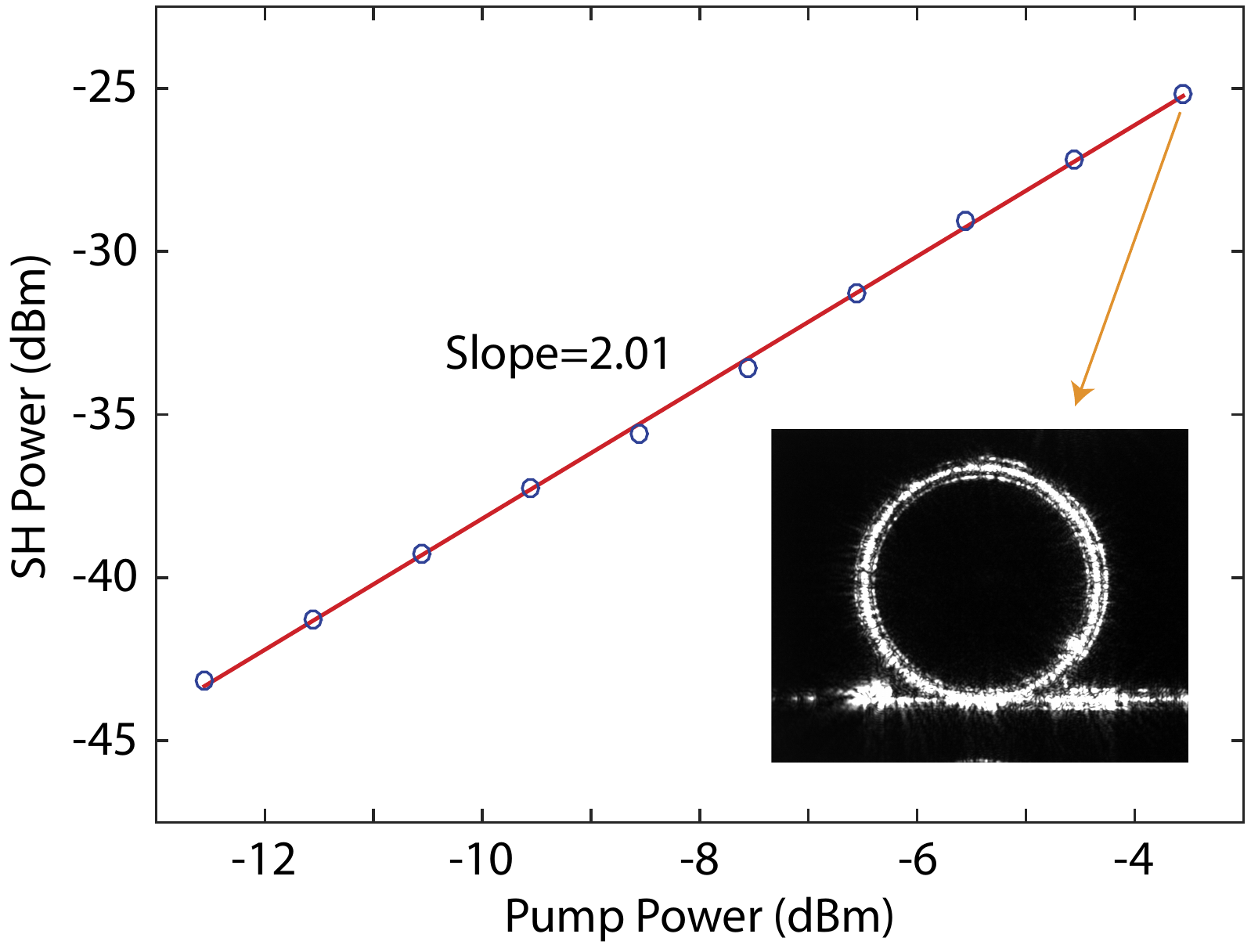}
	\caption{ Power dependence of SHG, showing a quadratic relation between the SH power and the pump power. The measured conversion efficiency is 1,500$\%~\rm {W^{-1}}$. The inset shows an optical image of generated SH light scattered from the microring when the pump power was 440 $\mu$W. } \label{Fig3}
\end{figure}

To study SHG in the microring resonator, we launched pump power into the FF mode at 1547.10 nm, and saw strong scattering of generated NIR light from the resonator by an optical microscope, with an example shown as the inset of Fig.~\ref{Fig3}. By varying the pump power, we obtained the power dependence of the SHG, as shown in Fig.~\ref{Fig3}. The experimental data exhibit a quadratic relation between the generated SH power and the FF pump power, which is the signature of SHG in the low-pump-power regime. The measured conversion efficiency is 1,500$\%~\rm{W^{-1}}$. This efficiency is more than one order of magnitude higher than those in many other LN microresonators \cite{Loncar14, Cheng16, Luo17OE, Xiao18, Zappe18, Huang18, Pertsch13, Jiang18, Li18}. It is even comparable with a recent study of cyclic phase matching in an X-cut microdisk exhibiting an ultra-high $Q$ of $\sim$10$^7$ \cite{Cheng18}, two orders of magnitude higher than that of our microring resonator, directly showing the advantage of exact modal phase matching. With future optimization of the optical $Q$'s of our device (say, by using a thicker LN film and an oxide cladding to reduce the sidewall scattering loss), we expect a further increase in the conversion efficiency.

\begin{figure}[t!]
	\centering\includegraphics[width=0.9\columnwidth]{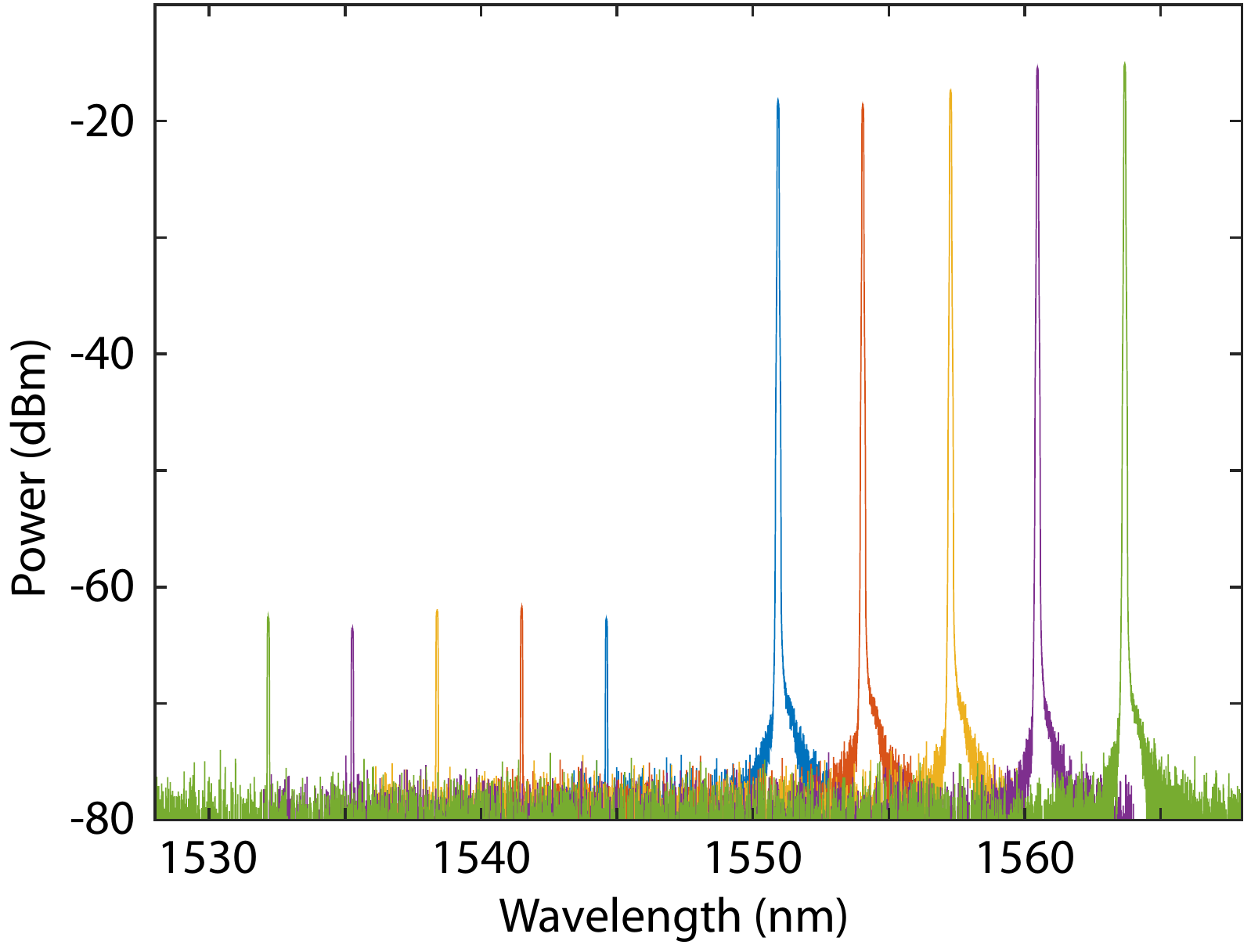}
	\caption{Recorded DFG spectra, when pumping at the SH mode in the NIR and one of the five nearest modes with longer wavelengths than the FF mode in the telecom band. Pump power at the SH mode was 6.6 $\mu$W.} \label{Fig4}
\end{figure}

The measured efficient SHG validated phase matching in our microring, and also indicated its capability of other parametric processes. In order to explore this, we launched power in both the SH mode, and one of the modes near the FF mode. Fig.~\ref{Fig4} presents the recorded spectra in the telecom band. With only 6.6 $\mu$W of on-chip power at the SH mode, we were able to convert long-wavelength telecom light coherently into shorter wavelengths through DFG. The long-wavelength pump power launched on chip was 105 $\mu$W, and the generated power at the difference frequencies was about 480 pW, indicating a conversion rate of about -53 dB.

\section{Theoretical analysis}
In order to acquire a better understanding of nonlinear parametric processes in our device, we can analyze the system with a model derived from the coupled mode theory \cite{BoydBook, Johnson07}. With no pump depletion, the conversion efficiency of modal-phase-matched SHG in our doubly resonant ($\omega_2=2\omega_1$) optical cavity can be calculated by
\begin{equation}
\Gamma \equiv \frac{P_2}{P_1^2 } \approx \frac{64 \gamma^2}{\hbar \omega_1^4} \frac{Q_{1l}^4 Q_{2l}^2}{Q_{1e}^2 Q_{2e}}, \label{Gamma}
\end{equation}
where $P_1$ ($P_2$) is the launched (produced) optical power at the FF (SH); $\hbar$ is the reduced Planck constant; $\omega_1$ ($\omega_2$) is the optical frequency, $Q_{1l}$ ($Q_{2l}$) is the loaded $Q$, and $Q_{1e}$ ($Q_{2e}$) is the coupling $Q$ of the FF (SH) mode; and $\gamma$ is the single-photon coupling strength written as
\begin{equation}
\gamma=\sqrt{\frac{\hbar \omega_1^2 \omega_2}{2\epsilon_0 \tilde{\epsilon}_1^2 \tilde{\epsilon}_2 }}\frac{d_{eff}\zeta}{\sqrt{V_{eff}}}. \label{gamma}
\end{equation}
In Eq.~\ref{gamma}, $\epsilon_0$ is the vacuum permittivity, $\tilde{\epsilon}_1$ ($\tilde{\epsilon}_2$) is the relative permittivity of the nonlinear medium at the FF (SH), $d_{eff}$ is the effective nonlinear coefficient, $\zeta$ is the mode overlap factor represented as 
\begin{equation}
\zeta = \frac{ \int_{\chi^{(2)}} (E_{1z}^*)^2 E_{2z} d^3x}{|\int_{\chi^{(2)}} |\vec{E}_1|^2 \vec{E}_1 d^3x|^{\frac{2}{3}} |\int_{\chi^{(2)}}  |\vec{E}_2|^2 \vec{E}_2 d^3x|^{\frac{1}{3}}}, \label{zeta}
\end{equation}
and $V_{eff} \equiv (V_{1}^2 V_{2})^{\frac{1}{3}}$ is the effective mode volume, with
\begin{equation}
V_{i} = \frac{ (\int_{ all} \epsilon_i |\vec{E}_i|^2 d^3x)^3 }{|\int_{\chi^{(2)}} \epsilon_i^{\frac{3}{2}} |\vec{E}_i|^2 \vec{E}_i d^3x|^2},(i=1,2),
\end{equation}
where  $\int_{\chi^{(2)}}$ and $\int_{all}$ denote three-dimensional integration over the nonlinear medium and all space, respectively, $\epsilon_1(\vec{r})$ [$\epsilon_2(\vec{r})$] is the relative permittivity at the FF (SH), and $E_{1z}$ ($E_{2z}$) is the z-component of $\vec{E}_1(\vec{r})$ [$\vec{E}_2(\vec{r})$], the electric field of the FF (SH) cavity mode.

With the equations above, the SHG efficiency in our LN microring is calculated to be $\Gamma \approx 30,000\%~\rm{W^{-1}}$. Thus, there is more than one order of magnitude difference between the theoretical prediction and our experimental result. The main reason for this discrepancy is likely non-uniformity of the microring at different azimuthal angles. By simulation, a change of 1 nm in the waveguide width, for example, will lead to a shift of $\sim-3$ nm in the phase-matched pump wavelength of SHG. Considering the small linewidths of our cavity modes, which are only 11 pm for the FF mode and 9 pm for the SH mode, the phase-matching window is easily shifted out of the cavity resonances due to fabrication imperfections. Besides, there is also a certain level of non-uniformity in the thickness of the LN thin film, giving more uncertainty to the uniformity of phase matching. We believe these technical issues will be addressed by optimized fabrication techniques in the near future, and the conversion efficiency can be significantly improved.

\section{Conclusion}

In conclusion, we have demonstrated optical parametric generation in a LN microring resonator with modal phase matching. We have used a single bus waveguide to conveniently couple the FF and SH modes, both showing coupling depths over 80\% and exhibiting loaded optical $Q$'s around $10^5$, resulting in a measured conversion efficiency of 1,500$\%~\rm{W^{-1}}$ for SHG. In addition, we have also observed DFG in the telecom band. Our work represents an important step towards ultra-highly efficient optical parametric generation in photonic circuits based on the LN integrated platform.

\section*{Acknowledgments}
The authors thank Xiyuan Lu at National Institute of Standards and Technology for helpful discussions. This work was supported in part by the National Science Foundation under Grant No.~ECCS-1641099, ECCS-1509749, and ECCS-1810169, and by the Defense Threat Reduction Agency under the Grant No.~HDTRA1827912. The project or effort depicted was or is sponsored by the Department of the Defense, Defense Threat Reduction Agency. The content of the information does not necessarily reflect the position or the policy of the federal government, and no official endorsement should be inferred. This work was performed in part at the Cornell NanoScale Facility, a member of the National Nanotechnology Coordinated Infrastructure (National Science Foundation, ECCS-1542081), and at the Cornell Center for Materials Research (National Science Foundation, DMR-1719875).

\bibliography{References}

\end{document}